\documentclass[12pt]{article}
\usepackage{amsfonts}
\usepackage[backend=bibtex,sorting=none,style=numeric-comp,maxnames=8]{biblatex}
\renewbibmacro{in:}{}

\usepackage[utf8]{inputenc} 
\usepackage{amsmath,amssymb} 
\usepackage{hyperref} 
\hypersetup{bookmarksopen=true, bookmarksnumbered=true,
 pdfstartview={FitH}, pdfborder={0 0 0}, colorlinks=true,
 linkcolor=red, linktocpage=true, citecolor=blue, urlcolor=blue,
 unicode=true, pdftitle={Conformal mass in EGB theory},
 pdfauthor={DJ}, pdfsubject={Einstein Gauss Bonnet theory,
   Kounterterms and AMD mass}, pdfkeywords={AdS, EGB, AMD,
   Kounterterms}}

\usepackage{geometry} 
\usepackage{fancyhdr} 
\usepackage{parskip} 
\usepackage[nottoc,notlof,notlot]{tocbibind} 
\usepackage[titles]{tocloft} 

\numberwithin{equation}{section} 

\bibliography{EGB}

\begin{document}
\title{\Huge\textbf{Conformal mass in Einstein-Gauss-Bonnet AdS gravity}}
\author{Dileep P. Jatkar$^1$\footnote{\href{mailto:dileep@hri.res.in}{dileep@hri.res.in}},
  Georgios Kofinas $^2$\footnote{\href{mailto:gkofinas@aegean.gr}{gkofinas@aegean.gr}},
 Olivera Miskovic$^3$\footnote{\href{mailto:olivera.miskovic@ucv.cl}{olivera.miskovic@ucv.cl}},
  Rodrigo Olea$^4$\footnote{\href{mailto:rodrigo.olea@unab.cl}{rodrigo.olea@unab.cl}}
  \bigskip\\
  \small{ $^1$Harish-Chandra Research Institute,} \\
  \small{Chhatnag Road, Jhunsi, Allahabad 211019, India}\\
  \small{$^2$Research Group of Geometry, Dynamical Systems and Cosmology}\\
  \small{Department of Information and Communication Systems Engineering}\\
  \small{University of the Aegean, Karlovassi 83200, Samos, Greece}\\
  \small{$^3$Instituto de F\' isica, Pontificia Universidad Cat\' olica de
    Valpara\' iso,}\\
  \small{Casilla 4059, Valpara\' iso, Chile}\\
  \small{ $^4$Departamento de Ciencias F\' isicas, }\\
  \small{Universidad Andres Bello, Rep\' ublica 220, Santiago, Chile}}

\maketitle
\thispagestyle{fancy}
\vspace{0.3cm}
\begin{abstract}
\normalsize

In this paper, we show that the physical information given by
conserved charges for asymptotically AdS spacetimes in
Einstein-Gauss-Bonnet AdS gravity is encoded in the electric part of
the Weyl tensor. This result generalizes the conformal mass definition
by Ashtekar-Magnon-Das (AMD) to a gravity theory with a Gauss-Bonnet
term. This proof makes use of the Noether charges obtained from an
action renormalized by the addition of counterterms which depend on
the extrinsic curvature (Kounterterms). If the asymptotic fall-off
behaviour of the Weyl tensor is same as the one considered in the AMD
method, then the Kounterterm charges and the AMD charges agree in any
dimension.
\end{abstract}
\newpage
\tableofcontents

\section{Introduction}

The AdS/CFT correspondence has been studied is great detail over the
years \cite{Maldacena:1997re}.  It is one of the most striking relations between a theory of
gravity and a theory without gravity \cite{Gubser:1998bc,Witten:1998qj,Aharony:1999ti}.  The central premise of this
correspondence is that all the information stored in the theory of
gravity is contained in the information stored in the boundary theory, and vice versa.  As
far as the theory of gravity is concerned, this fact was known in a
different guise much before the holographic relationship was proposed
in the AdS/CFT correspondence.  For example, it was known that one
cannot define local observables in the theory of gravity, which in
turn implies that one cannot define local energy-momentum tensor out of the metric.
The only sensible definition of the energy-momentum tensor is the
quasi-local energy-momentum tensor defined on the boundary.

Asymptotic symmetries and conserved charges do depend on the bulk
geometry and there are various ways one can define conserved charges
by studying asymptotic symmetries.  Among them, the Ashtekar-Magnon-Das \cite{Ashtekar:1984zz,Ashtekar:1999jx}
(AMD) method applies Penrose's conformal transformation to determine
conserved charges in the asymptotically AdS (AAdS) spaces. A defining feature of this method
is that all information about the charge is contained in the
electric part of the Weyl tensor,
\begin{equation}
\label{eq:elWeyl}
E^i_j=\frac{1}{D-3}\,W^{i\mu}_{j\nu}\,n_\mu n^\nu\,.
\end{equation}
Here, $n_\mu$ is the normal vector to the boundary, whose local coordinates are $x^i$.

In Einstein gravity, AMD charges are obtained by conformal completion techniques applied to AAdS spaces.
The conformal mapping between the physical  metric $g_{\mu\nu}$ (a given solution of the Einstein equations)
to an unphysical one, $\tilde{g}_{\mu\nu}=\Omega^2 g_{\mu\nu}$, serves the purpose of defining a regular boundary,
as long as the conformal factor $\Omega$ vanishes on the boundary and its derivative is finite. In doing so, the
conformal mass is encoded in the electric part of the Weyl tensor for $\tilde{g}_{\mu\nu}$. Suitable rescaling of
tensorial quantities make the charge formula expressible in terms of the tensor defined in Eq.(\ref{eq:elWeyl}),
instead.

In Ref.\cite{Jatkar:2014npa}, it was shown that in arbitrary
dimensions the AMD conserved charges for asymptotically AdS spaces
can be derived from the extrinsic curvature counterterms
(Kounter-terms \cite{Olea:2006vd,Kofinas:2006hr}) when the bulk gravity
action was given by the Einstein-Hilbert action.  In this gravity
theory, the on-shell Weyl tensor $W_{\alpha\beta}^{\mu\nu}$ is equal
to the AdS curvature tensor
$F_{\alpha\beta}^{\mu\nu}=R_{\alpha\beta}^{\mu\nu}+\frac{1}{\ell^2}\,\delta_{[\alpha\beta]}^{[\mu\nu]}$,
and the asymptotic behaviour of the Weyl tensor in the radial
direction is $W_{\alpha\beta}^{\mu\nu}=\mathcal{O}(1/r^{D-1})$. These
features of the theory in addition to the fact that the Kounterterm
charge factorizes over the AdS curvature projected to the boundary, $F_{kl}^{ij}$, are
the key ingredients in proving the equivalence of the Kounterterm mass
and the conformal mass.

In this paper, we will take up the problem of computing the conformal
mass in the Einstein-Gauss-Bonnet (EGB) AdS gravity using the
Kounterterm method in arbitrary dimensions.

In the EGB-AdS case, the on-shell Weyl tensor is not equal to the AdS curvature
tensor, {\em i.e.}
$W_{\alpha \beta }^{\mu \nu }\neq F_{\alpha \beta }^{\mu \nu }$.  The
difference, however, is subleading in $r$.  As the first step in
deriving the conformal mass, we will show that in the EGB theory the
difference between these two tensors is
$W_{\alpha \beta }^{\mu \nu}
=F_{\alpha \beta }^{\mu \nu}+\mathcal{O}(1/r^{2D-2})$.
This is true provided we define the AdS curvature
$F_{\alpha \beta }^{\mu \nu }$ in terms of the \textit{effective} AdS
radius.

We show the asymptotic relation between $W_{\alpha \beta }^{\mu \nu}$
and $F_{\alpha \beta }^{\mu \nu }$ only on a class of solutions. However, we expect that
it holds generically for localized configurations in AAdS spacetimes.
We then prove that the Kounterterm charges in EGB AdS gravity in all
dimensions also factorize by the boundary AdS curvature.
After these steps, the calculation of the conformal mass is the same
as in Einstein-Hilbert case. The obtained result matches the one of
Pang \cite{Pang:2011cs}, up to replacing the AdS radius by the effective one.

\section{Fall-off of the Weyl tensor in EGB AdS gravity}

Let us consider the $D$-dimensional spacetime $\mathcal{M} $ endowed with the metric $g_{\mu\nu}(x)$ whose
dynamics is described by the gravity action with the Einstein-Hilbert term,
the cosmological constant term and the Gauss-Bonnet term.  The
Gauss-Bonnet term is a specific quadratic combination of the Riemann
tensor $R_{\mu\nu\lambda\rho}$, the Ricci tensor $R_{\mu\nu}$ and the
Ricci scalar $R$,
\begin{equation}
  \label{eq:1}
  I_{\rm{EGB}} = \frac{1}{16\pi G} \int\limits_{\mathcal{M}} d^Dx \sqrt{-g}\left[R
  -2\Lambda
  +\alpha(R^2-4R_{\mu\nu}R^{\mu\nu}+R_{\mu\nu\lambda\rho}R^{\mu\nu\lambda\rho})
  \right]\ ,
\end{equation}
where the cosmological constant term is related to the AdS radius
$\ell$ by the relation
$\Lambda =-\frac{\left( D-1\right) \left( D-2\right) }{2\ell ^{2}}$
and $G$ is the $D$-dimensional gravitational constant.  The
Gauss-Bonnet coupling $\alpha$ is a dimensionful parameter with
spatial scaling dimension 2. If this parameter is related to the string
tension, it takes only nonegative values.

The Gauss-Bonnet term is important when the spacetime dimension
is $D\ge 5$. For $D=4$, the Gauss-Bonnet term is a topological invariant and
therefore does not contribute to the equations of motion.  It can,
however, modify the boundary dynamics owing to the fact that it is a
total derivative term in four dimensions.  In dimensions less than
four, the Gauss-Bonnet term vanishes identically.  For later
convenience, we will use the notation $\mathcal{L}$ for the Lagrangian
density to denote the term in the square brackets in (\ref{eq:1}).

The equation of motion derived from the action (\ref{eq:1}) takes the
form
\begin{equation}
  \label{eq:3}
  \mathcal{E}_{\nu }^{\mu }\equiv \, R_{\nu }^{\mu }-\frac{1}{2}\,R\,\delta _{\nu }^{\mu }+\Lambda \,\delta
_{\nu }^{\mu }+\alpha H_{\nu }^{\mu }=0\,,
\end{equation}
where the Lanczos tensor $H^\mu_\nu$ is the contribution coming from
the variation of the Gauss-Bonnet term.  This requires giving a
prescription for the variational principle for the manifolds with
boundary.  For the Einstein-Gauss-Bonnet theory we need to provide a
generalized Gibbons-Hawking term at the boundary for consistency of
the variational principle.  This has already been done in the
literature \cite{Myers:1987yn,Miskovic:2007mg}, what we will discuss in
the next section.  Since the Gauss-Bonnet term contains Riemann
tensor, it is useful to write variation of the Riemann tensor in terms
of variation of the Riemann-Christoffel connection
\begin{equation}
  \label{eq:6}
  \delta R^\mu_{\nu\lambda\rho} =
  \nabla_\lambda(\delta\Gamma^\mu_{\nu\rho}) -\nabla_\rho
  (\delta\Gamma^\mu_{\nu\lambda})\ .
\end{equation}
Using this expression for variation of the Riemann tensor, the Lanczos
tensor can be written as
\begin{equation}
  \label{eq:4}
  H_{\nu }^{\mu }=-\frac{1}{8}\,\delta_{[\nu \nu _{1}\cdots\nu_{4}]}
^{[\mu \mu _{1}\cdots \mu _{4}]}\,R_{\mu _{1}\mu _{2}}^{\nu _{1}\nu
_{2}}R_{\mu _{3}\mu _{4}}^{\nu _{3}\nu _{4}}\ .
\end{equation}
The tensor $\delta_{[\nu \nu _{1}\cdots \nu_{p}]}^{[\mu \mu _{1}\cdots \mu _{p}]}$
used in the definition of the Lanczos tensor in (\ref{eq:4}) is the
totally antisymmetric product of $p$ Kronecker delta symbols (see Appendix \ref{Delta}).

The AdS space is a solution to the Eq.(\ref{eq:3}), but with a
different than $\ell$ radius when $\alpha \neq 0$.  We will call it the effective AdS radius,
$\ell_{\rm{eff}}$, which can be derived from the following condition
\begin{equation}
  \label{eq:5}
  \frac{\alpha^*}{\ell_{\rm{eff}}^4}-\frac{1}{\ell_{\rm{eff}}^2}+\frac{1}{\ell^2}=0\,,\qquad \alpha^*=\alpha (D-3)(D-4)\, .
\end{equation}
Since the equation is quadratic in $\ell_{\rm{eff}}^2$, there are two AdS solutions with different effective radii when $0 \neq \alpha^* < \ell^2/4$.
The case where both AdS radii become the same
($\alpha^*= \ell^2/4$) represents an isolated
point of the theory, as the asymptotic behavior of the curvature is
radically different. In five dimensions, this particular value of the
GB coupling is referred to as the Chern-Simons point, with black holes
solutions which are almost of constant curvature \cite{Brihaye:2013vsa}. In
higher dimensions, it corresponds to a particular case of a large
family of theories known as Lovelock Unique Vacuum \cite{Kastor:2006vw}.

After manipulating the equation of motion (\ref{eq:3}) and denoting
$H=H^\mu_\mu$, we can write the following two on-shell conditions for
the Ricci tensor and the Ricci scalar,
\begin{eqnarray}
  \label{eq:7}
  R^\mu_\nu &=& \frac{\alpha}{D-2} \, H \delta^\mu_\nu - \alpha
  H^\mu_\nu-\frac{D-1}{\ell^2} \,\delta^\mu_\nu\,,\\
   \label{eq:8}
  R &=& \frac{2}{D-2}\left(\alpha H - \frac{D(D-1)(D-2)}{2\ell^2}\right)\ .
\end{eqnarray}
Using Eq.(\ref{eq:7}) and (\ref{eq:8}), the on-shell Weyl tensor can
be written as
\begin{eqnarray}
  \label{eq:9}
  W_{\alpha \beta }^{\mu \nu } &=&R_{\alpha \beta }^{\mu \nu }-\frac{1}{D-2}
  \,\delta _{[\alpha }^{[\mu }R_{\beta ]}^{\nu ]}+\frac{R}{(D-1)(D-2)}\,\delta _{[\alpha \beta]}^{[\mu \nu]} \nonumber\\
  &=&R_{\alpha \beta }^{\mu \nu }+\frac{1}{\ell ^{2}}\,\delta _{\alpha \beta}^{\mu \nu }
  +\frac{\alpha}{D-2}\,\left( \delta _{[\alpha }^{[\mu }H_{\beta]}^{\nu ]}
  -\frac{2}{D-1}\,H\,\delta _{[\alpha \beta] }^{[\mu \nu]}\right)\ ,
\end{eqnarray}
where
$\delta _{[ \alpha }^{[\mu }R_{\beta ]}^{\nu ]}=\delta _{\alpha
}^{\mu }R_{\beta }^{\nu }-\delta _{\beta }^{\mu }R_{\alpha }^{\nu
}-\delta _{\alpha }^{\nu }R_{\beta }^{\mu }+\delta _{\beta }^{\nu
}R_{\alpha }^{\mu }$.

The AdS space is a maximally symmetric solution to the
equations of motion of the Einstein-Hilbert action with negative
cosmological constant.  It is conformally flat space and, as a result,
the on-shell Weyl tensor $W^{\mu\nu}_{\rho\sigma}=R^{\mu\nu}_{\rho\sigma}+
\frac{1}{\ell^2}\,\delta^{[\mu\nu]}_{[\rho\sigma]}$
in this background vanishes. In what follows, we will refer to this tensor
as the AdS tensor with appropriate AdS radius.

In the Einstein-Gauss-Bonnet theory, however, the AdS solution is a
bit different, because it has a
different effective cosmological constant (equivalently, different AdS
radius).  As a result, the AdS tensor
$F_{\alpha \beta }^{\mu \nu }=R_{\alpha \beta }^{\mu \nu
}+\frac{1}{\ell _{\rm{eff}}^{2}}\,\delta _{\alpha \beta }^{\mu \nu }$
in the Einstein-Gauss-Bonnet theory differs from the Weyl tensor.  We
can write the Weyl tensor (\ref{eq:9}) as
\begin{eqnarray}
W_{\alpha \beta }^{\mu \nu } &=&R_{\alpha \beta }^{\mu \nu }
+\frac{1}{\ell _{\rm{eff}}^{2}}\,\delta _{\alpha \beta }^{\mu \nu }+X_{\alpha \beta }^{\mu\nu }\,, \label{eq:10}\\
X_{\alpha \beta }^{\mu \nu } &=&\frac{\alpha}{D-2}\,\delta _{[\alpha}^{[\mu }H_{\beta ]}^{\nu ]}
-\left( \frac{2\alpha\,H}{(D-1)(D-2)}+\frac{\alpha\,(D-3)(D-4)}{\ell _{\rm{eff}}^4}\right) \,
\delta_{[\alpha \beta] }^{[\mu \nu]}\,.
\end{eqnarray}
Clearly, when $\alpha =0$, the difference between the AdS curvature
and Weyl tensors vanishes as expected, $X_{\alpha\beta}^{\mu\nu} =0$.
A question is what is the behaviour of $X_{\alpha\beta}^{\mu\nu} $ for
$\alpha \neq 0$.

We will now use a heuristic argument to deduce a fall-off of the
tensor $X_{\alpha \beta }^{\mu \nu }$.  In the original discussion by
Ashtekar and Das \cite{Ashtekar:1999jx}, the fall-off of the Weyl
tensor in EH AdS gravity was based on dimensional analysis. Namely,
the Weyl tensor vanishes identically for the global AdS space, so the
difference between $\left. W\right|_{\rm{AdS}}$ and $W$ is due to a
non-vacuum state with total mass $M$, which should satisfy
$\left.W-W\right|_{\rm{AdS}}\sim GM$ asymptotically. Since the only
other dimensionful quantity is the radial coordinate, dimensional
analysis leads to the asymptotic relation
$W_{\alpha \beta }^{\mu \nu }\sim GM/r^{D-1}$.

On the other hand, the Weyl
tensor is on-shell equal to the AdS tensor in EH gravity,
$\left.X\right|_{\rm{EH}}=0$.
Thus, in EGB AdS gravity, the difference between the Weyl and AdS tensors is proportional to the GB coupling
and a quadratic mass correction of a non-vacuum state, that is
$\left. X-X\right|_{\rm{EH}}\sim \alpha \,(GM)^2$. This, in turn, implies the units
(length)$^{-2}$ of the tensor $X_{\alpha \beta }^{\mu \nu }$ when its
fall-off is of the form
\begin{equation}
X_{\alpha \beta }^{\mu \nu }\sim \alpha \,\frac{(GM)^2}{r^{2D-2}}\,.
\label{Xfall}
\end{equation}
We will prove that this asymptotic behavior is suitable for static black holes in EGB theory.

Let us consider the Einstein-Gauss-Bonnet AdS-Schwarzschild solution
\cite{Boulware:1985wk,Cai:2001dz}
\begin{eqnarray}
  \label{eq:11}
  ds^2 &=& -f(r) dt^2 + \frac{dr^2}{f(r)} + r^2
           \gamma_{mn}d\Omega^md\Omega^n\ ,  \nonumber\\
  f(r) &=& k+\frac{r^{2}}{2\alpha *}\left[ 1\pm
           \sqrt{1-4\alpha^*\left(\frac{1}{\ell^{2}}-\frac{\mu
           }{r^{D-1}}\right)}\right]\,,
\end{eqnarray}
where $\gamma_{mn}$ is the metric on a $(D-2)$-dimensional constant
curvature Riemannian space with $\Omega^m$ being the local coordinates,
and $k= 0,\ 1,\ {\rm or}\ -1$ represents flat, spherical and
hyperbolic metric on the $(D-2)$-dimensional space transverse to $r$
and $t$.  There are two branches of the theory and both of them have AdS asymptotics, with the effective radius
given by Eq.(\ref{eq:5}), that is,
\begin{equation}
\label{eq:leffpm}
\frac{1}{\ell_{\rm{eff}}^2}=\frac{1}{2\alpha^*}\left( 1\pm \sqrt{1-\frac{4\alpha^*}{\ell ^2}}\right) \,,  \qquad \alpha^*< \frac{\ell^2}{4}\,.
\end{equation}
Without loss of generality of the results, we shall choose one particular (non-degenerate) AdS vacuum with
radius $\ell_{\rm{eff}}$.

The Lanczos tensor can now be evaluated in this black hole background.
More details of the Lanczos tensor in this background are given in
Appendix \ref{lanczos-tensor}.  It is convenient to define a function
$P(r) =\left( \frac{f-k}{r^{2}}\right) ^{2}$, so that the Lanczos
tensor can be written in a simpler form.  In addition, it will be
helpful in the study of the asymptotic behaviour of the solution.  Let us
first rewrite the Lanczos tensor
(\ref{eq:12}), (\ref{eq:13}) in terms of $P(r)$,
\begin{eqnarray}
  \label{eq:14}
 H_r^r &=&H_t^t=-\frac{(D-2)(D-3)(D-4)}{2r^{D-2}}\,(r^{D-1}P)' \nonumber\\
  \label{eq:15}H_{m}^{n}
            &=&-\frac{(D-3)(D-4)}{2}\,\left[\rule{0pt}{16pt}r^{2}P^{\prime \prime }+2(D-1) rP'+(D-1)(D-2)P\right] \,\delta_m^n\ .
\end{eqnarray}
The function $P(r)$ evaluated on the solution reads
\begin{equation}
P =\frac{1}{4\alpha^{*2}}\left[ 1+\left( 1-\frac{2\alpha^*}{\ell_{\rm{eff}}^2}\right)^2
+\frac{4\alpha^*\mu }{r^{D-1}}\pm 2\sqrt{\left( 1-\frac{2\alpha^*}{\ell_{\rm{eff}}^2}\right)^2
+\frac{4\alpha^*\mu}{r^{D-1}}}\right].
\label{P}
\end{equation}
The asymptotic expansion of $P(r)$ will be convenient for analyzing
the asymptotic behavior of the solution. The small parameter of the
expansion is a dimensionless constant
$\varepsilon =4\alpha^*\mu /r^{D-1}<<1$. We note that the constant
$1-\frac{2\alpha^*}{\ell _{\rm{eff}}^{2}}$ appearing under the root
can be either positive or negative since, according to
(\ref{eq:leffpm}), we have
\begin{equation}
1-\frac{2\alpha^*}{\ell _{\rm{eff}}^{2}}=\mp \sqrt{1-\frac{4\alpha^*}{\ell ^{2}}}\,.
\end{equation}
Thus, the expansion will not depend on the choice of a particular branch. We obtain
\begin{eqnarray}
P &=&\frac{1}{\ell_{\rm{eff}}^{4}}+\frac{2\mu }{\left( 2\alpha^*-\ell _{\rm{eff}}^{2}\right) r^{D-1}}+\mathcal{O}(1/r^{2D-2})\,,
\nonumber   \\
rP' &=&-\frac{2(D-1)\mu }{\left( 2\alpha^*-\ell _{\rm{eff}}^{2}\right) r^{D-1}}+\mathcal{O}(1/r^{2D-2})\,,  \nonumber \\
r^{2}P^{\prime \prime } &=&\frac{2D(D-1)\mu }{\left( 2\alpha^*-\ell_{\rm{eff}}^{2}\right) r^{D-1}}+\mathcal{O}(1/r^{2D-2})\,,  \nonumber \\
\frac{(r^{D-1}P)^{\prime }}{r^{D-2}} &=&\frac{D-1}{\ell _{\mathrm{eff}}^{4}}+
\mathcal{O}(1/r^{2D-2})\,.
\end{eqnarray}
Using this leading behavior of $P(r)$, we find
that all mass terms cancel out in the asymptotic expansion and
the asymptotic behavior of the
Lanczos tensor evaluated on this solution is
\begin{equation}
\label{eq:47}
H_\mu^\nu = -\frac{(D-1)(D-2)(D-3)(D-4)}{2\ell_{\rm{eff}}^4}\,\delta_\mu^\nu\,+\mathcal{O}(1/r^{2D-2})\,.
\end{equation}
Substituting this behaviour of $H^\nu_\mu$ in
Eqs.(\ref{eq:10}), (\ref{eq:49}), a fall-off of the tensor
$X_{\alpha\beta}^{\mu\nu}$ that describes the difference between the
Weyl and the AdS curvature tensors can be evaluated as
\begin{equation}
\label{eq:48}
X_{\alpha\beta }^{\mu \nu }=\mathcal{O}(1/r^{2D-2})\,,
\end{equation}
in agreement with Eq.(\ref{Xfall}). Also note that the asymptotic
expansion of the function $P(r)$ defined in (\ref{P}) is in the small
quantity $4\alpha^*\mu/r^{D-1}$, that is $\alpha GM/r^{D-1}$, and the
first non-vanishing term is quadratic in mass, in accordance with
(\ref{Xfall}).

In the next section we will utilize this leading behavior of these
tensors to determine conserved charges in this theory.
We will derive this result using the Kounterterm
method.

\section{Conformal charge for EGB-AdS gravity}

In this section we determine the conserved charges in
Einstein-Gauss-Bonnet AdS gravity.  In order to derive equations of
motion, we need to define a variational principle for a theory of
gravity.  If the theory is defined on a manifold with boundary, we
need to impose boundary conditions on the metric so that the action
has an extremum on-shell.  The Dirichlet boundary condition on the
metric can be imposed by adding (generalized) Gibbons-Hawking terms to
the action which cancel variations of the extrinsic curvature coming
from the variation of the bulk action. In case of AdS-like geometries
where the boundary is defined along the radial direction, it is
convenient to write the metric in the Gauss-normal coordinates,
\begin{equation}
  \label{eq:16}
  ds^2 = g_{\mu\nu}dx^\mu dx^\nu=N^2(r)dr^2+h_{ij}(r,x)dx^i dx^j\, .
\end{equation}
The Gibbons-Hawking term is then written in terms of the extrinsic
curvature $K_{ij}(h)$ and the intrinsic curvature of the boundary
$\mathcal{R}^{ij}_{kl}(h)$.  The extrinsic curvature is given by the
Lie derivative of the induced metric $h_{ij}(r,x)$ on the boundary
along the outward pointing normal vector $n_\mu = (n_r, n_i) = (N, \vec 0)$,
\begin{equation}
  \label{eq:17}
  K_{ij} = -\frac12 \pounds_nh_{ij} = -\frac{1}{2N} \frac{\partial
    h_{ij}}{\partial r}\ .
\end{equation}
The intrinsic curvature of the boundary can be reexpressed in terms of
the bulk Riemann curvature tensor and the extrinsic curvature of the
boundary,
\begin{equation}
  \label{eq:18}
  \mathcal{R}^{ij}_{kl}(h) = R^{ij}_{kl}(g) +K^i_kK^j_l - K^i_lK^j_k\ .
\end{equation}
Then, the generalized Gibbons-Hawking term \cite{Myers:1987yn,Miskovic:2007mg}
for the Gauss-Bonnet theory has the form
\begin{eqnarray}
  \label{eq:19}
  I_{\rm{GGH}} &=& \int\limits_{\partial\mathcal{M}} d^{D-1}x \frac{\sqrt{-h}\,
  \delta_{[ i_{1}i_{2}i_{3}]}^{[j_{1}j_{2}j_{3}]}}{8\pi G}K_{j_{1}}^{i_{1}}\left[
\frac{\delta _{j_{2}}^{i_{2}}\delta _{j_{3}}^{i_{3}}}{(D-2)(D-3)}+2\alpha \,\left(
             \frac{1}{2}\,\mathcal{R}
_{j_{2}j_{3}}^{i_{2}i_{3}}-\frac{1}{3}\,K_{j_{2}}^{i_{2}}K_{j_{3}}^{i_{3}}%
\right) \right]\nonumber \\
&=&\!\!\!\! \int\limits_{\partial\mathcal{M}}\!\! d^{D-1}x \frac{\sqrt{-h}}{8\pi
    G}\,\left[ K+2\alpha K\left(
K^{ij}K_{ij}-\frac{1}{3}K^{2}\right) -\frac{2\alpha}{3}%
\,K_{k}^{i}K_{j}^{k}K_{i}^{j}-4\alpha\mathcal{G}^{ij}K_{ij} \right] ,
\end{eqnarray}
where $\mathcal{G}_{ij} = \mathcal{R}_{ij} - \frac{1}{2}\,\mathcal{R}\, h_{ij}$ is
the Einstein tensor derived from the induced metric $h_{ij}$ on the boundary.

In addition to the boundary Gibbons-Hawking term, we also need to
introduce local counterterms at the boundary to take care of
divergences.  In the AdS/CFT correspondence, these terms are derived
using holographic renormalization technique. However, this method
becomes cumbersome as one departs from the Einstein gravity and
includes higher-derivative terms.  There are various proposals to
handle higher-derivative terms, but all of them seem to work in
specific dimensions.  The Kounterterm regularization scheme
\cite{Olea:2006vd,Kofinas:2006hr,Miskovic:2010ui}, where the boundary term explicitly depends on the extrinsic curvature
$K_{ij}$, comes to rescue at this point.  In this
scheme, depending on even or odd dimensions, the counterterms are
either written in terms of topological terms, or in terms of the
Chern-Simons form in those dimensions.  In the Kounterterm
regularization method, the generalized Gibbons-Hawking term (\ref{eq:19}) is not added to the action, but the total
 action can be written as
\begin{equation}
  \label{eq:22}
  I_{\rm{ren}} = I_{\rm{EGB}} + c_{D-1}\int\limits_{\mathcal{\partial M}}d^{D-1}x\,
  B_{D-1}\ .
\end{equation}
Here, $c_{D-1}$ is a specific dimension dependent constant and
$B_{D-1}$ is the Kounterterm action which is written in terms of both
the boundary extrinsic curvature and the boundary intrinsic
curvature. The form of $B_{D-1}$ depends on whether $D$ is even or
odd.  The explicit form of the Kounterterm will be discussed later for
$D$ even, as well as $D$ odd.

Equation (\ref{eq:22}) gives a well-defined prescription for variation of
the metric.  Extremization of the total action is then given by
\begin{equation}
  \label{eq:23}
  \delta I_{\rm{ren}} = \int\limits_{\mathcal{\partial M}}d^{D-1}x \,(\Theta_{\rm{EGB}}
  +c_{D-1} \delta B_{D-1})\,,
\end{equation}
where $\Theta_{\rm{EGB}}$ is the boundary contribution coming from
variation of the action (\ref{eq:1}).

Let us now define conserved quantities.  They are associated with
certain global symmetries. In the theory of gravity, these
symmetries are asymptotic isometries of the background.  In our case we
will be interested in the diffeomorphism invariance,
\begin{equation}
  \label{eq:20}
  \delta_\epsilon g_{\mu\nu} = \pounds_\epsilon g_{\mu\nu}=-(\nabla_\mu
  \epsilon_\nu + \nabla_\nu \epsilon_\mu)\ .
\end{equation}
The variation of the total action $I_{\rm{ren}}$, which contains the
Einstein-Gauss-Bonnet action as well as the Kounterterm action, can be
succinctly written as
\begin{eqnarray}
  \label{eq:21}
  \delta_\epsilon I_{\rm{ren}} &=&\int\limits_{\partial \mathcal{M}}d^{D-1}x\,n_{\mu }\left[ \frac{1}{N}
\,\Theta ^{\mu }(\epsilon)+\sqrt{-h}\,\epsilon^{\mu }\mathcal{L}+
c_{D-1}\,n^{\mu}\,\partial _{i}\left( \epsilon^{i}B_{D-1}\right) \right]   \notag \\
&&-\frac{1}{16\pi G}\int\limits_{\mathcal{M}}d^{D}x\,\sqrt{-g}\,
\mathcal{E}^{\mu \nu }\pounds _{\epsilon }g_{\mu \nu}\ ,
\end{eqnarray}
where $n_\mu\Theta^\mu$ is the boundary contribution coming from the
variation of the Einstein-Gauss-Bonnet action.  It is now easy to read
the conserved current from the variation of the action,
\begin{equation}
  \label{eq:24}
  J^\mu(\epsilon) = \Theta ^{\mu }(\epsilon)+\sqrt{-g}\,
  \epsilon^{\mu }\mathcal{L}+
  c_{D-1}\,N\, n^{\mu}\,\partial _{i}\left( \epsilon^{i}B_{D-1}\right)\ .
\end{equation}
The conserved charge $Q[\epsilon]$ is in fact the boundary term
appearing in the variation (\ref{eq:21}),
\begin{equation}
  \label{eq:25}
  Q[\epsilon] = \int\limits_{\partial \mathcal{M}}d^{D-1}x\,
  \frac{1}{N} \,n_{\mu}J^\mu (\epsilon) \ .
\end{equation}
The conserved charge is typically computed by integrating the density
over a codimension two hypersurface.  The Noether charge defined above
is not an integral over a codimension two space and, in general, it is
not obvious that the boundary integral will reduce to an integral over
a codimension one subspace of the boundary.  In the case of
Einstein-Gauss-Bonnet AdS theory, the conserved current projected
along the outward pointing normal can be written as a total
derivative.  In the Gauss normal coordinates, the boundary integral
can be further restricted to a codimension one subspace $\Sigma$ of
the boundary.  The radial component of the conserved current can be
written as
\begin{equation}
  \label{eq:27}
  J^r = \frac{1}{N} \,n_{\mu}J^\mu=\partial_j\left(\sqrt{-h}\epsilon^i(q^j_i +
    q^j_{(0)i})\right)\ ,
\end{equation}
where $q^j_{(0)i}$ is the contribution of vacuum energy which is
non-vanishing in odd dimensions only.  The charge density
tensor $q^j_i$, in general, contains information about all the charges
that a black hole can carry.  In our case, however, this charge would
be the mass of the black hole.  This charge vanishes when evaluated in
vacuum.  Using the fact that $J^r$ can be written as a total
derivative, we can re-express the Noether charge as an integral over a
codimension one subspace of the boundary,
\begin{equation}
  \label{eq:28}
  Q[\epsilon] = \int\limits_\Sigma d^{D-2} y \sqrt{\sigma}
  u_j\epsilon^i \left( q^j_i + q^j_{(0)i}\right)\ ,
\end{equation}
where $u_i$ is the unit time-like normal to the boundary of the
spatial section described by the metric $\sigma$.

For global AdS space, in particular $F_{kl}^{ij}=0$, the charge
density $q_{i}^{j}$ vanishes so that $q_{(0)i}^{j}$ can be interpreted
as the vacuum energy tensor of the spacetime (which cannot be
  observed in perturbative, background-dependent methods in EGB theory
  \cite{Deser:2002jk,Deruelle:2003ps,Petrov:2011vz}).  The fact that
$q_{i}^{j}=0$ in the vacuum suggests that $q_{i}^{j}$ should be
factorizable by $F$, making this vanishing explicit. Indeed, denoting
$n=[D/2]$, it can be shown (see Appendix \ref{sec:even-dimensions} and
\ref{sec:odd-dimensions}) that the tensor $q_{i}^{j}$ can always be
written as
\begin{equation}
q_{i}^{j}=a_{n}\,\delta _{[i_{1}\cdots i_{2n-1}]}^{[jj_{2}\cdots
j_{2n-1}]}\,K_{i}^{i_{1}}F_{j_{2}j_{3}}^{i_{2}i_{3}}\mathcal{J}_{j_{4}\cdots
j_{2n-1}}^{i_{4}\cdots i_{2n-1}}(F)\,,  \label{q_general}
\end{equation}
where $a_n$ is a constant that depends on dimension and the GB coupling,
and $\mathcal{J}_{j_4\cdots j_{2n-1}}^{i_4\cdots i_{2n-1}}(F)$ is a given
polynomial of the AdS curvature in integral representation.

To study the conserved charges further, we will have to consider two
cases: (\emph{i}) $D$ is even and (\emph{ii}) $D$ is odd.  This distinction is required
for two reasons.  The conserved charges are derived from the
Kounterterms, and latter have one form in even dimensions and a different
form in odd dimensions.  This implies that the form of the charge density
tensor also differs between odd and even dimensions.  Second, the
term $q^j_{(0)i}$ vanishes in even dimensions but is nonzero in odd
dimensions.  As a result, it is convenient to split the discussion and
study even-dimensional and odd-dimensional cases separately.

\subsection{Even dimensions, $D=2n$}

The relevant Kounterterms in even dimensions are \cite{Miskovic:2010ui,Kofinas:2006hr}
\begin{eqnarray}
  \label{eq:29}
  B_{2n-1} &=&2n\sqrt{-h}\int\limits_{0}^{1}dt\,\delta _{\lbrack i_{1}\cdots
i_{2n-1}]}^{[j_{1}\cdots j_{2n-1}]}\,K_{j_{1}}^{i_{1}}\left( \frac{1}{2}\,
\mathcal{R}_{j_{2}j_{3}}^{i_{2}i_{3}}
-t^{2}K_{j_{2}}^{i_{2}}K_{j_{3}}^{i_{3}}\right)\times   \notag \\
&&\qquad \cdots \times \left( \frac{1}{2}\,
\mathcal{R}_{j_{2n-2}j_{2n-1}}^{i_{2n-2}i_{2n-1}}-t^{2}K_{j_{2n-2}}^{i_{2n-2}}K_{j_{2n-1}}^{i_{2n-1}}\right) \,,
\end{eqnarray}
and the coupling constant in (\ref{eq:22}), fixed by the variational principle, is
\begin{equation}
c_{2n-1}=-\frac{(-\ell_{\rm{eff}}^2)^{n-1}}{16\pi G\,n(2n-2)!}\,.\label{c2n-1}
\end{equation}

The regularized charge density tensor (\ref{eq:28}) in even dimensions has the form
\begin{eqnarray}
\label{eq:38}
q_{i}^{j} &=&\frac{1}{16\pi G\left( 2n-2\right) !2^{n-2}}\,\delta _{\lbrack
i_{1}\cdots i_{2n-1}]}^{[jj_{2}\cdots j_{2n-1}]}\,K_{i}^{i_{1}}\times
\nonumber \\
&&\times \left[ \rule{0in}{0.24in}\left( \delta _{\lbrack
j_{2}j_{3}]}^{[i_{2}i_{3}]}+2\alpha \left( 2n-2\right) \left( 2n-3\right)
R_{j_{2}j_{3}}^{i_{2}i_{3}}\right) \delta _{\lbrack
j_{4}j_{5}]}^{[i_{4}i_{5}]}\cdots \delta _{\lbrack
j_{2n-2}j_{2n-1}]}^{[i_{2n-2}i_{2n-1}]}\right.   \nonumber \\
&&\left. \rule{0in}{0.24in}-\left( -\ell _{\rm{eff}}^{2}\right)
^{n-1}\left( 1-\frac{2\alpha }{\ell _{\rm{eff}}^{2}}\,\left( 2n-2\right)
\left( 2n-3\right) \right) R_{j_{2}j_{3}}^{i_{2}i_{3}}\cdots
R_{j_{2n-2}j_{2n-1}}^{i_{2n-2}i_{2n-1}}\right] \,.
\end{eqnarray}
As discussed before, the tensor $q^j_{(0)i}$ vanishes. The tensor $q^j_i$
can be factorized by $F^{ij}_{kl}$ in the form (\ref{q_general}), in
terms of an integral (\ref{eq:44}) involving only the AdS curvature
tensor with the constant $a_n=-\frac{nc_{2n-1}}{ 2^{n-2}}$. The
details of this calculation are given in Appendix
\ref{sec:even-dimensions}.

Recall that the Weyl tensor decreases asymptotically as
\begin{equation}
\label{fall-off}
W_{\alpha\beta}^{\mu\nu}=\mathcal{O}(1/r^{D-1})\,.
\end{equation}
Using the behaviour of $X_{\alpha\beta}^{\mu\nu}$ in Eq.(\ref{eq:48})
and substituting it in the Eq.(\ref{eq:10}), we see that
$F_{kl}^{ij}=W_{kl}^{ij}+\mathcal{O}(1/r^{2D-2})$.  This allows us to
trade the AdS tensor $F_{kl}^{ij}$ off in terms of the Weyl tensor
$W_{kl}^{ij}$ in Eq.(\ref{q_general}) because, according to
(\ref{fall-off}), it decreases milder than $r^{2D-2}$ for large
distances. From Eq.(\ref{eq:17}), it is easy to see that
\begin{equation}
K_{j}^{i}=-\frac{1}{\ell _{\rm{eff}}}\,\delta _{j}^{i}+\mathcal{O}(1/r^2)\,.
\end{equation}
On account of the fall-off of the Weyl tensor and the behaviour of $K^i_j$, the expression of
$q^i_j$ is equivalent (up to a relevant order) to
\begin{equation}
q_{i}^{j}=-\frac{\left( -\ell _{\rm{eff}}^{2}\right) ^{n-1}}{16\pi G
\ell _{\rm{eff}}\left( 2n-2\right) !2^{n-2}}\,\delta _{\lbrack ii_{2}\cdots
i_{2n-1}]}^{[jj_{2}\cdots j_{2n-1}]}\,W_{j_{2}j_{3}}^{i_{2}i_{3}}
\mathcal{J}_{j_{4}\cdots j_{2n-1}}^{i_{4}\cdots i_{2n-1}}(0)\,.
\end{equation}
We have substituted the AdS tensor $F$ in the expression of
$\mathcal{J}$ by zero because of its fast fall-off and the fact that higher powers of the Weyl tensor do not
contribute to the conserved charge.

Setting $F=0$ in
Eq.(\ref{eq:44}), we can evaluate the integral because it essentially
becomes trivial, $\int_{0}^{1}du=1$,
\begin{eqnarray}
\mathcal{J}(0) &=&\frac{\gamma \left( n-2\right)
-\left( n-1\right) }{\left( -\ell _{\rm{eff}}^{2}\right) ^{n-2}}\,\left(
\delta ^{\lbrack 2]}\right) ^{n-2}  \nonumber \\
&=&-\frac{n-1}{\left( -\ell _{\rm{eff}}^{2}\right) ^{n-2}}\,\left( 1-\frac{%
2\alpha ^{\ast }}{\ell _{\rm{eff}}^{2}}\right) \left( \delta ^{\lbrack
2]}\right) ^{n-2}\,.
\end{eqnarray}
As a result of this simplification, the
charge density tensor also simplifies to
\begin{equation}
q_{i}^{j} =-\frac{\ell _{\rm{eff}}\left( n-1\right) }{16\pi G\,\left(
2n-2\right) !2^{n-2}}\,\left( 1-\frac{2\alpha ^{\ast }}{\ell _{\rm{eff}%
}^{2}}\right) \delta _{\lbrack ii_{2}\cdots i_{2n-1}]}^{[jj_{2}\cdots
j_{2n-1}]}\,W_{j_{2}j_{3}}^{i_{2}i_{3}}\,\delta _{\lbrack
j_{4}j_{5}]}^{[i_{4}i_{5}]}\cdots \delta _{\lbrack
j_{2n-2}j_{2n-1}]}^{[i_{2n-2}i_{2n-1}]} \label{eq:53}\,.
\end{equation}
The last expression can be calculated using the
identity (\ref{eq:52}) given in Appendix \ref{Delta}.
For simplification of Eq.(\ref{eq:53}) the relevant values of the variables
are $N=2n-1$, $k=2n-4$ and $p=2n-1$.  This gives the relation
\begin{equation}
\delta _{[ ii_{2}\cdots i_{2n-1}]}^{[jj_{2}\cdots j_{2n-1}]}\,\delta
_{j_{4}}^{i_{4}}\cdots \delta _{j_{2n-1}}^{i_{2n-1}}=(2n-4)!\,
\delta _{\left[ ii_{2}i_{3}\right] }^{\left[ jj_{2}j_{3}\right] }\,.
\end{equation}
This further simplifies the expression of the charge density tensor
and allows us to write it in a compact form,
\begin{equation}
\label{eq:26}
q_{i}^{j}=-\frac{\ell _{\rm{eff}}}{32\pi G\,\left( 2n-3\right) }\,\left( 1-%
\frac{2\alpha ^{\ast }}{\ell _{\rm{eff}}^{2}}\right) \delta _{\left[
ii_{2}i_{3}\right] }^{\left[ jj_{2}j_{3}\right] }
\,W_{j_{2}j_{3}}^{i_{2}i_{3}}\,.
\end{equation}
Expansion of the Kronecker delta implies $\delta _{\left[ ii_{2}i_{3}\right]
}^{\left[ jj_{2}j_{3}\right] }\,W_{j_{2}j_{3}}^{i_{2}i_{3}}=2\delta
_{i}^{j}\,\big( W_{kl}^{kl}-2W_{ki}^{kj}\big) $. Furthermore, since the Weyl
tensor is traceless ($W_{\mu \beta }^{\mu \alpha }=0$), it enables to
write the charge density tensor in terms of the electric part of the Weyl tensor,
that is
\begin{equation}
q_{i}^{j}=-\frac{\ell _{\text{eff}}}{8\pi G\,\left( 2n-3\right) }\,\left( 1-
\frac{2\alpha ^{\ast }}{\ell _{\text{eff}}^{2}}\right) \,W_{ri}^{rj}\,.
\end{equation}

Recall that in the even spacetime dimensions, the tensor
$q_{(0)i}^{j}$ vanishes identically, $q_{(0)i}^{j}=0$.  We can then use the result
(\ref{eq:26}) and substitute it in Eq.(\ref{eq:28}), which in a
straightforward manner (as in \cite{Jatkar:2014npa}) leads to
\begin{equation}
Q_{\rm{EGB}}[\epsilon ]=-\frac{\ell _{\rm{eff}}}{8\pi G\left( D-3\right) }
\left( 1-\frac{2\alpha ^{\ast }}{\ell_{\rm{eff}}^2}\right)
\int\limits_{\Sigma _{\infty }}d\Sigma \,W_{ir}^{jr}\,\epsilon ^{i}u_{j}\,.
\label{conformal charge}
\end{equation}

Our results are in agreement with those of \cite{Pang:2011cs}, where the conformal mass was calculated in general quadratic
curvature gravity using a different method. Restricting the coupling constants to EGB gravity in AdS space with the radius $\ell_{\rm{eff}}$,
this charge matches Eq.(\ref{conformal charge}).

\subsection{Odd dimensions, $D=2n+1$}

The Kounterterms in odd dimensions are \cite{Miskovic:2010ui,Kofinas:2006hr}
\begin{eqnarray}
  \label{eq:37}
  B_{2n} &=&2n\sqrt{-h}\int\limits_{0}^{1}dt\int\limits_{0}^{t}ds\,\delta
_{\lbrack i_{1}\cdots i_{2n}]}^{[j_{1}\cdots
j_{2n}]}\,K_{j_{1}}^{i_{1}}\delta _{j_{2}}^{i_{2}}\left( \frac{1}{2}\,%
\mathcal{R}%
_{j_{3}j_{4}}^{i_{3}i_{4}}-t^{2}K_{j_{3}}^{i_{3}}K_{j_{4}}^{i_{4}}+\frac{%
s^{2}}{\ell_{\rm{eff}}^2}\,\delta _{j_{3}}^{i_{3}}\delta
_{j_{4}}^{i_{4}}\right) \times  \notag \\
&&\cdots \times \left( \frac{1}{2}\,\mathcal{R}%
_{j_{2n-1}j_{2n}}^{i_{2n-1}i_{2n}}-t^{2}K_{j_{2n-1}}^{i_{2n-1}}K_{j_{2n}}^{i_{2n}}+%
\frac{s^{2}}{\ell_{\rm{eff}}^2}\,\delta _{j_{2n-1}}^{i_{2n-1}}\delta
_{j_{2n}}^{i_{2n}}\right) \,.
\end{eqnarray}
The value of the coupling $c_{2n}$ which appears in the action (\ref{eq:22}) is
\begin{equation}
c_{2n}=-\frac{1}{16\pi G}\frac{(-\ell _{\rm{eff}}^2)^{n-1}}{2^{2n-2}n(n-1)!^2}
\left( 1-\frac{2\alpha }{\ell _{\rm{eff}}^2}\,(2n-1)(2n-2) \right) \,.  \label{c_2n}
\end{equation}
The expression for the charge density tensor $q^i_j$, similar to the
even-dimensional one (\ref{eq:38}), is given by Eq.(\ref{eq:54}) in
Appendix \ref{sec:odd-dimensions}.  This appendix also contains
various technical details and identities required to simplify the form
of $q^i_j$.

After an integral parametrization of the constant $c_{2n}$ and factorization of $q^i_j$ by $F^{ij}_{kl}$,
the charge density tensor $ q_{i}^{j}$ takes the
form (\ref{q_general}), with the constant  $a_n=-\frac{nc_{2n-1}}{(1-\gamma)\, 2^{n-2}}$,
where $\gamma =\frac{2\alpha}{\ell_{\rm{eff}}^2}\,(2n-1)(2n-2)$.

For completeness, let us write the expression of $q^i_j$ with all indices in
place,
\begin{equation}
  \label{eq:69}
  q_{i}^{j} =\frac{n\,c_{2n}\,\ell _{\rm{eff}}}{\ell _{\rm{eff}}^{2(n-1)}\left( 1-\gamma \right) }\,\delta _{[ ii_{2}i_{3}i_{4}\cdots
i_{2n}]}^{[jj_{2}j_{3}i_{4}\cdots i_{2n}]}\,F_{j_{2}j_{3}}^{i_{2}i_{3}}
\mathcal{J}_{j_4\cdots j_{2n}}^{i_4\cdots i_{2n}}(F)\,.
\end{equation}

As in the even-dimensional case, we use the asymptotic behavior of the
Weyl tensor (\ref{fall-off}) and the relation
$F_{kl}^{ij}=W_{kl}^{ij}+\mathcal{O}(1/r^{2D-2})$ to conclude that the
AdS tensor and Weyl tensor have the same fall-off at the leading
order, and
\begin{equation}
\label{eq:68}
K_{j}^{i}=-\frac{1}{\ell _{\rm{eff}}}\,\delta _{j}^{i}+\mathcal{O}(1/r^2)\,.
\end{equation}
This allows us to replace $F$ by $W$ in Eq.(\ref{eq:69}).  The tensor
$\mathcal{J}$ depends on $F$, but since higher
powers of $F$ would fall off faster, they do not contribute to the
conserved charge.  We will therefore substitute $F=0$ in the
expression of $\mathcal{J}$ and the explicit AdS tensor in the
expression of $q^i_j$ will be replaced by the Weyl tensor $W$. Thus,
\begin{equation}
  \label{eq:70}
  q_{i}^{j} =\frac{n\,c_{2n}\,\ell _{\rm{eff}}}{\ell _{\rm{eff}}^{2(n-1)}(1-\gamma)}
  \,\delta _{\lbrack ii_{2}i_{3}i_{4}\cdots i_{2n}]}^{[jj_{2}j_{3}i_{4}\cdots i_{2n}]}\,W_{j_{2}j_{3}}^{i_{2}i_{3}}
\mathcal{J}_{j_4\cdots j_{2n}}^{i_4\cdots i_{2n}}(0)\,.
\end{equation}
The integral $\mathcal{J}$, after setting $F=0$, reads
\begin{equation}
\label{eq:72}
\mathcal{J}(0)=\int\limits_{0}^{1}du\,\left[ -\left( n-1\right) \left(
u^{2}-1\right) ^{n-2}+\gamma \,u^{2\left( n-1\right) }-\left( n-1\right)
\left( n-2\right) \gamma \int\limits_{0}^{1}ds\,s\,\left( u^{2}-s\right)
^{n-3}\right] .
\end{equation}
We can carry out the integral involving the variable $s$ using
\begin{equation}
\label{eq:73}
\int\limits_{0}^{1}ds\,s\,\left( u^{2}-s\right) ^{n-3}=\frac{%
u^{2(n-1)}-\left( u^{2}-1\right) ^{n-2}\left( u^{2}+n-2\right) }{\left(
n-1\right) \left( n-2\right) }\,.
\end{equation}
Substituting Eq.(\ref{eq:73}) in Eq.(\ref{eq:72}), we get
\begin{equation}
\label{eq:74}
\mathcal{J}(0)=\int\limits_{0}^{1}du\,\left[ -\left( n-1\right) +\gamma \left(
u^{2}+n-2\right) \right] \left( u^{2}-1\right) ^{n-2}\,.
\end{equation}
It is now easy to use the identity Eq.(\ref{eq:71}) to write
\begin{equation}
\label{eq:75}
\mathcal{J}(0)=\left( -1\right) ^{n-1}\,\frac{\left( n-1\right) !\left(
n-2\right) !\,2^{2n-4}}{\left( 2n-3\right) !}\left( 1-\gamma \,\frac{2n-3}{2n-1}\right) \,.
\end{equation}

This form of $\mathcal{J}(0)$ simplifies the expression of the
charge density tensor, which can now be written as
\begin{equation}
\label{eq:76}
q_{i}^{j}=-\frac{\ell _{\rm{eff}}}{32\pi G(2n-2)}
\left( 1-\frac{2\alpha ^{\ast }}{\ell _{\rm{eff}}^{2}}\right) \delta _{\left[
ii_{2}i_{3}\right] }^{\left[ jj_{2}j_{3}\right] }
\,W_{j_{2}j_{3}}^{i_{2}i_{3}}\,.
\end{equation}

In the end, we see that both in even dimensions (\ref{eq:26}) and odd
dimensions (\ref{eq:76}), the expression of the charge density tensor is
of the form
\begin{equation}
\label{eq:77}
q_{i}^{j}=-\frac{\ell _{\rm{eff}}}{32\pi G\,\left( D-3\right) }\,\left( 1-%
\frac{2\alpha ^{\ast }}{\ell _{\rm{eff}}^{2}}\right) \delta _{\left[
ii_{2}i_{3}\right] }^{\left[ jj_{2}j_{3}\right] }%
\,W_{j_{2}j_{3}}^{i_{2}i_{3}}\,.
\end{equation}
Thus, the conserved charge $Q_{\rm EGB}[\xi]$ in odd dimensional is also
given by the expression (\ref{conformal charge}).

\section{Conclusions}

In this paper, we have shown that the information regarding mass and
other conserved quantities for AAdS solutions in EGB theory is always
contained in the electric part of the Weyl tensor.  This comparison
has been carried out by direct expansion of the charge formulas
obtained within the Kounterterm regularization for AdS gravity, just
assuming a standard asymptotic fall-off of the metric.

The asymptotic charges are derived using the surface terms in the
variation of the action, which includes boundary terms. A key role is
played by the appearance of the Weyl tensor in this surface term, only
for a fine-tuned coupling $c_{d}$ given by Eq.(\ref{c2n-1}) in even
dimensions and Eq.(\ref{eq:55}) in odd dimensions.

This form of the surface term does not lend itself to a clear
definition of the quasilocal stress tensor, because the counterterms
added here are extrinsic and not purely intrinsic, as in the standard
methods in EGB theory \cite{Brihaye:2008xu,Liu:2008zf}.  However,
the general expression of the variation of the action can shed some
light on the problem of separability of the quasilocal stress tensor
in odd dimensions.  In particular, the quasilocal stress
tensor should contain a piece which is written in terms of the
electric part of the Weyl tensor and the rest should give rise to
the terms which contribute to the vacuum energy for the AAdS space,
 as in Eq.(\ref{eq:28}).

Since the expression for the surface term is the same regardless
of the type of variation performed, any transformation of the
fields will preserve the vacuum configuration, as the Weyl tensor
vanishes identically for maximally-symmetric spacetimes. At the same
time, at least in even dimensions, the value of the action is
identically zero for global AdS.

In the Kounterterm regularization method, the action is
stationary under arbitrary variation of the fields, including the
contributions that appear at the boundary, provided we impose
specific asymptotic conditions on the curvature.  This method can be
used to study supersymmetric extensions of the AdS gravity action.
The closure of supersymmetry not only in the bulk, but also at the
boundary, might be the ultimate reason behind the success of the
Kounterterm regularization method.  In this regard, there is new
evidence supporting this claim given in
Ref.\cite{Andrianopoli:2014aqa} that in four dimensions, demanding the
vanishing of (super) AdS curvature at the boundary, one is able to fix
the coupling of Gauss-Bonnet term.  Recall that the addition of
  the GB term is equivalent to Holographic Renormalization in 4D AdS
  gravity \cite{Miskovic:2009bm}.  It would be interesting to use this
  to extend the relation between renormalized AdS action and
  supersymmetry \cite{Andrianopoli:2014aqa} to higher dimensions.

\section*{Acknowledgements}

This work was funded by FONDECYT Grants
  No.~1131075 and No.~1110102. O.M. thanks DII-PUCV for support
  through the project No.~123.711. The work of R.O. is financed
  in part by the UNAB grant No.~DI-551-14/R.  The work of D.P.J.
  is partly supported by DAE project 12-R\& D-HRI-5.02-0303.

\appendix

\section{Kronecker delta of rank $p$ \ \label{Delta}}
In this appendix we will write down the notation for the rank $p$
Kroneckar delta.  This notation is taken from Ref.\cite{Miskovic:2010ui}.
The totally-antisymmetric Kronecker delta of rank $p$ is defined as the
determinant
\begin{equation}
\label{eq:39}
\delta _{\left[ \mu _{1}\cdots \mu _{p}\right] }^{\left[ \nu _{1}\cdots \nu
_{p}\right] }:=\left\vert
\begin{array}{cccc}
\delta _{\mu _{1}}^{\nu _{1}} & \delta _{\mu _{1}}^{\nu _{2}} & \cdots &
\delta _{\mu _{1}}^{\nu _{p}} \\
\delta _{\mu _{2}}^{\nu _{1}} & \delta _{\mu _{2}}^{\nu _{2}} &  & \delta
_{\mu _{2}}^{\nu _{p}} \\
\vdots &  & \ddots &  \\
\delta _{\mu _{p}}^{\nu _{1}} & \delta _{\mu _{p}}^{\nu _{2}} & \cdots &
\delta _{\mu _{p}}^{\nu _{p}}%
\end{array}%
\right\vert \,.
\end{equation}
A contraction of $k\leq p$ indices in the Kronecker delta of rank $p$
produces a delta of rank $p-k$,
\begin{equation}
\delta _{\left[ \mu _{1}\cdots \mu _{k}\cdots \mu _{p}\right] }^{\left[ \nu
_{1}\cdots \nu _{k}\cdots \nu _{p}\right] }\,\delta _{\nu _{1}}^{\mu
_{1}}\cdots \delta _{\nu _{k}}^{\mu _{k}}=\frac{\left( N-p+k\right) !}{%
\left( N-p\right) !}\,\delta _{\left[ \mu _{k+1}\cdots \mu _{p}\right] }^{%
\left[ \nu _{k+1}\cdots \nu _{p}\right] }\,,
\end{equation}%
where $N$ is the range of indices.

If $N$ is the range of indices, a contraction of $k$ indices in the
Kronecker delta of rank $p$ produces a delta of rank $p-k$,
\begin{equation}
\label{eq:52}
\delta _{\left[ \mu _{1}\cdots \mu _{k}\cdots \mu _{p}\right] }^{\left[ \nu
_{1}\cdots \nu _{k}\cdots \nu _{p}\right] }\,\delta _{\nu _{1}}^{\mu
_{1}}\cdots \delta _{\nu _{k}}^{\mu _{k}}=\frac{\left( N-p+k\right) !}{%
\left( N-p\right) !}\,\delta _{\left[ \mu _{k+1}\cdots \mu _{p}\right] }^{%
\left[ \nu _{k+1}\cdots \nu _{p}\right] }\,.
\end{equation}%

\section{The Lanczos Tensor in EGB AdS Schwarzschild background \label{lanczos-tensor}}

Consider the Einstein-Gauss-Bonnet AdS-Schwarzschild solution
\cite{Boulware:1985wk}
\begin{eqnarray}
  \label{eq:46}
  ds^2 &=& -f(r) dt^2 + \frac{dr^2}{f(r)} + r^2 \gamma_{mn}d\Omega^md\Omega^n\, , \nonumber \\
  f(r) &=& k+\frac{r^{2}}{2\alpha *}\left[ 1\pm \sqrt{1-4\alpha^*\left(\frac{1}{\ell^{2}}-\frac{\mu }{r^{D-1}}\right)}
  \right]\,.
\end{eqnarray}
The Lanczos tensor (\ref{eq:4}) for this solution is given by
\begin{eqnarray}
  \label{eq:12}
 H_{r}^{r} &=&H_{t}^{t}=-(D-2)(D-3)(D-4) \,\frac{f-k}{r^{2}}
  \,\left( \frac{f^{\prime }}{r}+\frac{D-5}{2}\,\frac{f-k}{r^{2}}\right), \\
H_{m}^{n} &=&-(D-3)(D-4)\,\delta _{m}^{n}\,\left[
  \frac{f-k}{r^{2}}\,f^{\prime \prime }+\left( \frac{f^{\prime}}{r}\right) ^{2}
  +2\left(D-5\right) \,\frac{f-k}{r^{2}}\frac{f^{\prime }}{r}\right. \nonumber\\
  &&+\left. \frac{1}{2}\,\left( D-5\right) \left( D-6\right) \,\left( \frac{f-k%
  }{r^{2}}\right) ^{2}\right]\ .
  \label{eq:13}
\end{eqnarray}
The Lanczos tensor can be defined in terms of an auxiliary function $P(r) =
\left(\frac{f-k}{r^2}\right)^2$.  The components of the tensor
$X^{\mu\nu}_{\alpha\beta}$ can be written in terms of the components
of the Lanczos tensor as
\begin{eqnarray}
  \label{eq:49}
  X_{tr}^{tr} &=&\frac{2\alpha}{D-2}\, H_{r}^{r}
  -\frac{2\alpha H}{(D-1)(D-2)}-\frac{\alpha^*}{\ell_{\rm{eff}}^{4}}\, \nonumber \\
  X_{\alpha\beta}^{\mu\nu} &=& X_{tr}^{tr}\delta^{[\mu\nu]}_{[\alpha\beta]}
  =\left(\frac{1}{4\alpha^*}-\frac{1}{\ell_{\rm{eff}}^2}\right)\,
  \delta_{[\alpha \beta]}^{[\mu \nu]}\, .
\end{eqnarray}

\section{Useful identities in EGB AdS gravity \label{Identities}}

Here we present the detailed derivation of the charge density tensor in
even and odd dimensions.  We manipulate them to write in a
simpler form which is quoted in the main text.

\subsection{Even dimensions}
\label{sec:even-dimensions}

In $D=2n$ dimensions, the regularized charge density tensor is given by the
formula (\ref{eq:38}), which can be explicitly written down using the definition of
the rank $p$ Kronecker delta given in Eq.(\ref{eq:39}),
\begin{eqnarray}
\label{eq:40}
q_{i}^{j} &=&\frac{1}{16\pi G\left( 2n-2\right) !2^{n-2}}\,\delta
^{j[2n-1]}\,K_{i}\left[ \rule{0in}{0.24in}\left( \delta ^{\lbrack
2]}+2\alpha \left( 2n-2\right) \left( 2n-3\right) R\right) \left( \delta
^{\lbrack 2]}\right) ^{n-2}\right.   \nonumber \\
&&\left. -\left( -\ell _{\rm{eff}}^{2}\right) ^{n-1}\left( 1-\frac{2\alpha
}{\ell _{\rm{eff}}^{2}}\,\left( 2n-2\right) \left( 2n-3\right) \right)
R^{n-1}\right] \,,
\end{eqnarray}%
where $\delta^{[2]}$ is a shorthand notation for
$\delta^{[ij]}_{[kl]}$ and similarly $\delta^{j[2n-1]}$ is the
Kronecker delta of rank $2n-1$ with $j$ being the only unpaired index.
This expression of $q_{i}^{j}$ can further be simplified by defining
\begin{equation}
\label{eq:41}
\Delta =\frac{1}{\ell _{\rm{eff}}^{2}}\,\delta ^{\lbrack 2]}\,,\qquad
\gamma =\frac{2\alpha }{\ell _{\rm{eff}}^{2}}\,\left( 2n-2\right) \left(
2n-3\right) \,.
\end{equation}%
Using Eq.(\ref{eq:41}), we can write the charge density tensor given in
Eq.(\ref{eq:40}) as
\begin{equation}
q_{i}^{j}=\frac{\left( -\ell _{\rm{eff}}^{2}\right) ^{n-1}}{16\pi G\left(
2n-2\right) !2^{n-2}}\,\delta ^{j[2n-1]}\,K_{i}\left\{ \rule{0in}{0.24in}%
\gamma R\left[ R^{n-2}-\left( -\Delta \right) ^{n-2}\right] -\left[
R^{n-1}-\left( -\Delta \right) ^{n-1}\right] \right\} \,.
\end{equation}%
This form of the charge density tensor can now be manipulated using
the following identity,
\begin{equation}
  \label{eq:42}
  R^{n-1}-\left( -\Delta \right) ^{n-1}=\left( n-1\right) \left( R+\Delta
\right) \int\limits_{0}^{1}du\,\left[ u\left( R+\Delta \right) -\Delta %
\right] ^{n-2}\, .
\end{equation}
This identity allows us to write $q_{i}^{j}$ as an integral formula,
\begin{equation}
  \label{eq:43}
  q_{i}^{j}=\frac{\left( -\ell _{\rm{eff}}^{2}\right) ^{n-1}}{16\pi G
  (2n-2! \,2^{n-2}}\,\delta ^{j[2n-1]}\,K_{i}\left( R+\Delta \right)
\mathcal{J}(R+\Delta)\,,
\end{equation}
where the function $\mathcal{J}(R+\Delta)$ contains the
integral over the variable $u$,
\begin{equation}
  \label{eq:44}
  \mathcal{J}(R+\Delta )=\int\limits_{0}^{1}du\,\left\{ \rule%
{0in}{0.24in}\gamma R\left( n-2\right) \left[ u\left( R+\Delta \right)
-\Delta \right] ^{n-3}-\left( n-1\right) \left[ u\left( R+\Delta \right)
-\Delta \right] ^{n-2}\right\} \,.
\end{equation}
Although it is not obvious from the form of the integral, the quantity
$\mathcal{J}(R+\Delta)$ is an antisymmetric tensor with
$2n-4$ covariant and $2n-4$ contravariant indices.  To see this
explicitly, let us substitute the form of $\Delta$ in the
expression of $\mathcal{J}(R+\Delta)$.  Since $R+\Delta $ is
exactly the AdS curvature $F_{kl}^{ij}=R_{kl}^{ij}+\frac{1}{\ell
  _{\rm{eff}}^{2}}\,\delta_{\lbrack kl]}^{[ij]}$,
the charge density tensor $q_i^j$  can be written as
\begin{equation}
  \label{eq:45}
  q_{i}^{j}=\frac{\left( -\ell _{\rm{eff}}^{2}\right) ^{n-1}}{16\pi G\left(
2n-2\right) !2^{n-2}}\,\delta _{\lbrack i_{1}\cdots
i_{2n-1}]}^{[jj_{2}\cdots j_{2n-1}]}\,K_{i}^{i_{1}}
F_{j_{2}j_{3}}^{i_{2}i_{3}} \mathcal{J}_{j_{4}\cdots
j_{2n-1}}^{i_{4}\cdots i_{2n-1}}(F)\,.
\end{equation}

\subsection{Odd dimensions}
\label{sec:odd-dimensions}

In odd $D=2n+1$ dimensions, the formula for the charge density tensor reads
\cite{Miskovic:2010ui,Kofinas:2006hr}
\begin{eqnarray}
\label{eq:54}
q_{i}^{j} &=&\frac{1}{16\pi G\left( 2n-1\right) !2^{n-2}}\,\delta _{\lbrack
i_{1}\cdots i_{2n}]}^{[jj_{2}\cdots j_{2n}]}\,K_{i}^{i_{1}}\delta
_{j_{2}}^{i_{2}}\times   \nonumber \\
&&\times \left[ \rule{0in}{0.26in}\left( \delta _{\lbrack
j_{3}j_{4}]}^{[i_{3}i_{4}]}+2\alpha \left( 2n-1\right) \left( 2n-2\right)
\,R_{j_{3}j_{4}}^{i_{3}i_{4}}\right) \delta _{\lbrack
j_{5}j_{6}]}^{[i_{5}i_{6}]}\cdots \delta _{\lbrack
j_{2n-1}j_{2n}]}^{[i_{2n-1}i_{2n}]}\right.   \nonumber \\
&&\hspace{-1cm}+\left. 16\pi G\left( 2n-1\right)
!\,nc_{2n}\int\limits_{0}^{1}du\left( R_{j_{3}j_{4}}^{i_{3}i_{4}}+\frac{u^{2}%
}{\ell _{\rm{eff}}^{2}}\,\delta _{\lbrack
j_{3}j_{4}]}^{[i_{3}i_{4}]}\right) \cdots \left(
R_{j_{2n-1}j_{2n}}^{i_{2n-1}i_{2n}}+\frac{u^{2}}{\ell _{\rm{eff}}^{2}}
\,\delta _{[j_{2n-1}j_{2n}]}^{[i_{2n-1}i_{2n}]}\right) \right]
\end{eqnarray}
where the constant $c_{2n}$ has the form (\ref{c_2n}), or equivalently
\begin{equation}
\label{eq:55}
c_{2n}=-\frac{1}{16\pi G}\frac{\ell _{\rm{eff}}^{2\left( n-1\right) }}{%
n\left( 2n-1\right) !}\left( 1-\frac{2\alpha }{\ell _{\rm{eff}}^{2}}%
\,\left( 2n-1\right) \left( 2n-2\right) \right) \left[ \int\limits_{0}^{1}du%
\,\left( u^{2}-1\right) ^{n-1}\right] ^{-1}.
\end{equation}
The equivalence between (\ref{c_2n}) and (\ref{eq:55}) is explicit after evaluating the integral,
\begin{equation}
\label{eq:71}
\int\limits_{0}^{1}du\,\left( u^{2}-1\right) ^{n-1}= (-1)^{n-1}\, \frac{2^{2n-2}(n-1)!^2}{(2n-1)!}\,.
\end{equation}

The charge (\ref{eq:54}) can be written in a more compact form by
suppressing repeated indices,
\begin{eqnarray}
\label{eq:56}
q_{i}^{j} &=&\frac{1}{16\pi G(2n-1)!2^{n-2}}\,\delta^{j[2n-1]}\,K_{i}\,
\delta \left[ \rule{0in}{0.46in}\left(\delta ^{[2]}+2\alpha (2n-1)(2n-2) \,R\right)(\delta^{[2]})^{n-2}\right.   \nonumber \\
&&+\left. 16\pi G (2n-1)!\,nc_{2n}\int\limits_{0}^{1}du
\left( R+\frac{u^{2}}{\ell _{\rm{eff}}^{2}}\,\delta ^{\lbrack 2]}\right) ^{n-1}\right].
\end{eqnarray}
This compact notation is the same as the one used in Appendix
\ref{sec:even-dimensions}.
However, instead of carrying out this integral, we will pull out the
constant $c_{2n}$ so that the $u$-integral is introduced in the first
line as well.  With this rearrangement, the charge density tensor can
be written is a compact form as an integral
\begin{equation}
\label{eq:57}
q_{i}^{j}=-\frac{nc_{2n}}{2^{n-2}\left( 1-\gamma \right) }\,\delta
^{j[2n-1]}\,K_{i}\,\delta \int\limits_{0}^{1}du\,\mathcal{Q}(R,u)\,,
\end{equation}%
where we have used the same notation as in the even-dimensional case
but with $2n$ replaced by $2n+1$,
\begin{equation}
\label{eq:58}
\Delta =\frac{1}{\ell _{\rm{eff}}^{2}}\,\delta ^{\lbrack 2]}\,,\qquad
\gamma =\frac{2\alpha }{\ell _{\rm{eff}}^{2}}\,\left( 2n-1\right) \left(
2n-2\right) \,.
\end{equation}%
The integrand is given by
\begin{eqnarray}
\label{eq:59}
\mathcal{Q}(R,u) &=&\left( u^{2}-1\right) ^{n-1}\Delta ^{n-1}-\left(
R+u^{2}\Delta \right) ^{n-1}  \nonumber \\
&&+\gamma \,\left[ R\Delta ^{n-2}\left( u^{2}-1\right) ^{n-1}+\left(
R+u^{2}\Delta \right) ^{n-1}\right] \,.
\end{eqnarray}%
As in the previous subsection \ref{sec:even-dimensions}, we will
repeatedly use the formula (\ref{eq:42}) to factorize the expression
of $\mathcal{Q}$ so that a factor of $\Delta +R$ can be pulled out.

Let us consider the first line of Eq.(\ref{eq:59}).  Using the formula
(\ref{eq:42}), we get
\begin{equation}
\label{eq:64}
\left( u^{2}-1\right) ^{n-1}\Delta ^{n-1}-\left( R+u^{2}\Delta \right)
^{n-1}=-\left( n-1\right) \left( R+\Delta \right) \int\limits_{0}^{1}ds\,
\left[ \left( u^{2}-s\right) \Delta +\left( 1-s\right) R\right] ^{n-2}.
\end{equation}%
The second line of $\mathcal{Q}$ in Eq.(\ref{eq:59}) needs repeated
application of the formula (\ref{eq:42}).  Before we use Eq.(\ref{eq:42}),
 let us first write the term as
\begin{equation}
\label{eq:60}
\left( R+u^{2}\Delta \right) ^{n-1}+R\Delta
^{n-2}\left( u^{2}-1\right) ^{n-1}=\left( R+\Delta \right) \left( \Delta
^{n-2}u^{2\left( n-1\right) }+Y\right) \,,
\end{equation}%
where the tensor $Y$ is determined by demanding that it satisfies
\begin{equation}
\label{eq:61}
\left( R+\Delta \right) Y=\left( R+u^{2}\Delta \right) ^{n-1}-\left(
u^{2}\Delta \right) ^{n-1}+R\Delta ^{n-2}\left[ \left( u^{2}-1\right)
^{n-1}-\left( u^{2}\right) ^{n-1}\right] \,.
\end{equation}
To extract the form of $Y$, we will manipulate the right-hand side of
(\ref{eq:61}).  This is done by applying Eq.(\ref{eq:42}),
which results in
\begin{equation}
\label{eq:62}
\left( R+\Delta \right) Y=\left( n-1\right) R\int\limits_{0}^{1}ds\,\left[
\left( sR+u^{2}\Delta \right) ^{n-2}-\Delta ^{n-2}\left( u^{2}-s\right)
^{n-2}\right] \,.
\end{equation}%
The expression on the right of (\ref{eq:62}) is still not convenient
for us to read out the formula for $Y$.  We will apply the same
formula for the second time to the expression in the square brackets
in (\ref{eq:62}).  This gives us the desired result.  After stripping
off the $R+\Delta$ factor, we obtain%
\begin{equation}
\label{eq:63}
Y=\left( n-1\right) \left( n-2\right)
R\int\limits_{0}^{1}ds\,s\int\limits_{0}^{1}dt\,\left[ tsR+\left(
u^{2}-s+ts\right) \Delta \right] ^{n-3}\,.
\end{equation}%
Using Eqs.(\ref{eq:64}), (\ref{eq:63}), we can finally write the integrand
$\mathcal{Q}(R,u)$ in the factorized form,
\begin{eqnarray}
\mathcal{Q}(R,u)\!\! &=&\!\!\left( R+\Delta \right) \left\{ -\left(
n-1\right) \int\limits_{0}^{1}ds\,\left[ \left[ \left( u^{2}-s\right) \Delta
+\left( 1-s\right) R\right] ^{n-2}\right] +\gamma \,\Delta ^{n-2}u^{2\left(
n-1\right) }\right.   \nonumber \\
&&\left. +\left( n-1\right) \left( n-2\right) \gamma
R\int\limits_{0}^{1}ds\,s\int\limits_{0}^{1}dt\,\left[ tsR+\left(
u^{2}-s+ts\right) \Delta \right] ^{n-3}\right\} \,.
\end{eqnarray}%
Substituting this expression of $\mathcal{Q}(R,u)$ in the
integral representation of the charge density tensor, we get
\begin{eqnarray}
\label{q exact}
q_{i}^{j} &=&\!\!\frac{-nc_{2n}}{2^{n-2}\left( 1-\gamma \right) }\,\delta
^{j[2n-1]}\,K_{i}\,\left( R+\Delta \right) \delta
              \int\limits_{0}^{1}du\,\left[ \left(1- n\right)
\int\limits_{0}^{1}ds\,\left( \frac{u^{2}-s}{\ell _{\rm{eff}}^{2}}\,\delta
^{\lbrack 2]}+\left( 1-s\right) R\right) ^{n-2}\right.   \nonumber \\
&&+\frac{\gamma }{\ell _{\rm{eff}}^{2(n-2)}}\,\left( \delta ^{\lbrack
2]}\right)^{n-2}\ u^{2\left( n-1\right) }  \nonumber \\
&&\left. +\left( n-1\right) \left( n-2\right) \gamma
R\int\limits_{0}^{1}ds\,s\int\limits_{0}^{1}dt\,\left( tsR+\frac{u^{2}-s+ts}{%
\ell _{\rm{eff}}^{2}}\,\delta ^{\lbrack 2]}\right) ^{n-3}\right] \,.
\end{eqnarray}%
Like in the even-dimensional case we will write the charge density
tensor in a compact form using
\begin{eqnarray}
  \label{eq:65}
 \mathcal{J}(R+\Delta) &=&  -\left(
n-1\right) \int\limits_{0}^{1}ds\,\left[ \left[ \left( u^{2}-s\right) \Delta
+\left( 1-s\right) R\right] ^{n-2}\right] +\gamma \,\Delta ^{n-2}u^{2\left(
n-1\right) }  \nonumber \\
&&\!\!\! +\left( n-1\right) \left( n-2\right) \gamma
R\int\limits_{0}^{1}ds\,s\int\limits_{0}^{1}dt\,\left[ tsR+\left(
u^{2}-s+ts\right) \Delta \right] ^{n-3}\, .
\end{eqnarray}
Here, to simplify the notation, we have suppressed indices on
$\mathcal{J}$.  As mentioned in the previous subsection, $R+\Delta$ is
equal to the AdS curvature $F_{kl}^{ij}=R_{kl}^{ij}+\frac{1}{
  \ell_{\rm{eff}}^{2}}\,\delta_{[kl]}^{[ij]}$.  We can
therefore replace $R+\Delta$ in the charge density expression by the
AdS tensor $F$,
\begin{equation}
  \label{eq:66}
q_{i}^{j}=\frac{nc_{2n}\ell _{\rm{eff}}}{2^{n-2}\ell _{\rm{eff}%
}^{2(n-1)}\left( 1-\gamma \right) }\,\delta ^{j[2n-1]}\,\delta _{i}\,\delta
\left( \delta ^{\lbrack 2]}\right) ^{n-2}F\mathcal{J}(F)\,,
\end{equation}
where again we have suppressed repeated indices on the Kronecker deltas.

\printbibliography

\end{document}